\documentclass[fdp,a4paper,fleqn%
]{w-art}
\usepackage{times,cite,w-thm}
\theoremstyle{plain}

\theoremstyle{definition}

\usepackage[]{graphicx}
\begin{document}
\DOIsuffix{theDOIsuffix}
\Volume{55}
\Month{01}
\Year{2007}
\pagespan{1}{}
\Receiveddate{XXXX}
\Reviseddate{XXXX}
\Accepteddate{XXXX}
\Dateposted{XXXX}



\title[Neutrino physics]{Neutrino physics beyond neutrino masses}


\author[F. del Aguila]{F. del Aguila%
  \footnote{E-mail:~\textsf{faguila@ugr.es} 
}}
\address[]{CAFPE and Departamento de F\'{\i}sica Te\'orica y del 
Cosmos, Universidad de Granada, E-18071 Granada, Spain}
\author[J. de Blas]{J. de Blas\footnote{E-mail:~\textsf{deblasm@ugr.es}}}
\author[A. Carmona]{A. Carmona\footnote{E-mail:~\textsf{adrian@ugr.es}}}
\author[J. Santiago]{J. Santiago\footnote{E-mail:~\textsf{jsantiago@ugr.es}}}
\begin{abstract}
We briefly summarise the current status of neutrino masses
and mixing, paying special attention to the prospects for 
observing new leptonic interactions. 
\end{abstract}
\maketitle                   





\section{Introduction}

The observed neutrino properties are in agreement with the 
minimal extension of the Standard Model (SM) resulting from 
the sole addition of neutrino masses \cite{Amsler:2008zzb,Mohapatra:1998rq}.  
Neutrinos are massless within the SM because there are not neutrino 
singlet counterparts, Higgs fields transform as an electroweak doublet, 
and the theory is renormalizable in the old sense. 
However, neutrino masses are generated relaxing any of these 
three conditions. 
 Indeed, the addition of three right-handed (RH) neutrinos $\nu_R$ 
allows for arbitrary Dirac neutrino masses 
after electroweak symmetry breaking, $<\phi^0> = v/\sqrt 2$, 
\begin{equation}
  \label{eq:Diracneutrinomasses}
  - y^\nu_{\alpha \beta} \overline{l_L^\alpha} \tilde \phi \nu_R^\beta 
+ {\rm h.c.} \rightarrow - y^\nu_{\alpha \beta} \frac{v}{\sqrt 2} 
\overline{\nu_L^\alpha} \nu_R^\beta + {\rm h.c.}\ ,
\hspace{0.4cm}
{\rm where}\ l = \left(\begin{array}{c} \nu  \\ \ell \end{array}\right)
\ {\rm and}\ \tilde\phi = i \sigma_2 \phi^* \ ,  
\end{equation}
similarly as for up quarks but with Yukawa couplings $y^\nu_{\alpha \beta}$ 
much smaller. 
This lepton number conserving (LNC) term and the 
corresponding neutrino mass matrix then provide the observed neutrino 
masses and charged current mixing upon diagonalisation, 
\begin{equation}
  \label{eq:Diracdiagonalization}
  m_{\nu i} = U^\dagger_{L i\alpha} y^\nu_{\alpha \beta} 
\frac{v}{\sqrt 2} U_{R \beta i} \ ,
\hspace{0.4cm}
{\rm with}\ U_{L,R} \ {\rm unitary \ matrices ,\ and}
\end{equation}
\begin{equation}
  \label{eq:Chargedcurrent}
  {\cal L}_W = - \frac{g}{\sqrt 2}\ \overline{\ell_L^\alpha} 
\gamma ^\mu U_{L \alpha i} \nu_L^i W^{-}_\mu + {\rm h.c.} \ . 
\end{equation}
Neutrino oscillations, which is the only manifestation of 
neutrino masses and mixing up to now, are well described 
by two mass splittings $\Delta m^2_{ij} \equiv m^2_{\nu i} - m^2_{\nu j}$ 
and the PMNS mixing matrix \cite{Pontecorvo:1957cp}, 
which equals $U_L$ when $l_L^\alpha$ are 
written in the charged lepton mass eigenstate basis. 
A global fit to available data gives 
\cite{GonzalezGarcia:2007ib} 
\begin{equation}
  \label{eq:limitsonmassesandmixing}
\begin{array}{c} \Delta m^2_{21} =  
7.67^{+0.67}_{-0.61} \times 10^{-5} {\rm eV}^2, \\ 
\Delta m^2_{31} = \left\{ \begin{array}{c}  
\hspace{-0.2cm}-2.37^{+0.43}_{-0.46} \times 10^{-3} {\rm eV}^2, \\ 
\hspace{-0.2cm}+2.46^{+0.47}_{-0.42} \times 10^{-3} {\rm eV}^2, \end{array} \right.
\end{array} 
\hspace{-0.1cm} 
|U_L| = \left(\begin{array}{ccc} 
0.77-0.86 & 0.50-0.63 & 0.00-0.22 \\ 
0.22-0.56 & 0.44-0.73 & 0.57-0.80 \\ 
0.21-0.55 & 0.40-0.71 & 0.59-0.82 \end{array}\right).
\end{equation}
The variation ranges correspond to 3$\sigma$ errors, 
and the negative and positive 
$\Delta m^2_{31}$ figures stand for inverted and normal hierarchy, 
respectively. 
There are excellent reviews on neutrino oscillation 
experiments \cite{Oscilaciones} and global fits 
\cite{Fits}, which are in good agreement within the 
available precision. 
(We ignore LSND data \cite{Aguilar:2001ty}.)
Note that these experiments do not distinguish between 
Dirac and Majorana masses, because in both cases the neutrino 
charged gauge interactions are given by Eq. (\ref{eq:Chargedcurrent}),  
and neutral gauge interactions also only involve 
left-handed (LH) neutrinos and are universal at lowest order 
\cite{delAguila:2007ug} (see also \cite{NSI}).  
In this scenario and in the absence of other light fields 
further new physics can be parameterized by the corresponding 
effective Lagrangian. Current limits on non-standard 
operators \cite{NSIFits,delAguila:2009vv} are presented in next section. 

With the minimal SM fermion content neutrinos can have 
Majorana masses if we add Higgs triplets or 
we allow for higher order operators. This second case in particular  
includes the first one when the Higgs triplet is integrated out. 
At any rate there is lepton number violation (LNV), and 
neutrino masses result from the famous dimension five 
Weinberg operator \cite{Weinberg:1979sa} ${\cal O}_5$ 
after electroweak symmetry breaking
\begin{equation} 
\frac{x_{\alpha \beta}}{\Lambda} {\cal O}_5^{\alpha \beta} = 
\frac{x_{\alpha \beta}}{\Lambda} \overline{(l_L^\alpha)^c}\tilde \phi ^* 
\tilde \phi^\dagger l_L^\beta  
\rightarrow \frac{x_{\alpha \beta}}{\Lambda} 
\frac{v^2}{2}\overline{(\nu^\alpha _L)^c} \nu^\beta _L \ , 
\hspace{0.2cm} 
m_{\nu i} = - U^\dagger_{L i\alpha} \frac{x^\dagger_{\alpha \beta}}{\Lambda} 
v^2 U^*_{L \beta i} \ . 
\label{Op5}  
\end{equation} 
In this case the tiny neutrino masses are due to the 
very small operator coefficients $x_{\alpha \beta}/\Lambda$ 
multiplying ${\cal O}_5^{\alpha \beta}$, 
which are so minuscule because $x_{\alpha \beta}$ are extremely small 
($\sim 10^{-12}$ for $\Lambda \sim v$) 
or $\Lambda$ is very large 
($\sim 10^{14}$ for $x_{\alpha \beta} \sim 1$). 
In the latter case, as $\Lambda$ 
is the effective scale in the Lagrangian expansion, all higher order 
effects from higher dimensional operators are negligible. 
Hence, the phenomenologically relevant situation at LHC 
is the first one, with new physics near the TeV scale 
and the dimensionless operator coefficients quite small on 
other grounds. In this set-up there are two possibilities, 
too: that the see-saw messengers generating ${\cal O}_5$ 
are near the electroweak scale, $M_{SSM} = \Lambda$, or that 
the new physics near this scale does not mediate the 
see-saw mechanism, $M_{SSM} \gg \Lambda$ but is 
related to it within a given model. 
We will discuss both cases in Section \ref{Majorananeutrinos}. 
As already emphasized, Majorana masses give the same 
neutrino oscillation predictions as Dirac ones in this minimal 
SM extension with only ${\cal O}_5$. But in this 
case the neutrino mass matrix is symmetric and 
$U_{R \beta i}$ in Eq. (\ref{eq:Diracdiagonalization}), 
which plays no r\^ole in neutrino oscillations, is equal 
to $U^*_{L \beta i}$ (see Eq. (\ref{Op5})).  

Independently of the neutrino mass character any realistic model 
must reproduce the observed spectrum. 
Although many models can accommodate the values in 
Eq. (\ref{eq:limitsonmassesandmixing}), there is no 
compelling, simple and predictive theory of lepton 
flavor. But, as we emphasize in Section \ref{Models}, 
in contrast with the quark sector lepton mixing is 
rather close to tri-bi-maximal mixing \cite{Harrison:2002er}, what 
seems to indicate that a flavor symmetry slightly 
broken is at work. 


\section{Current limits on new neutrino interactions}
\label{Limitsonneutrinointeractions}
   
Neutrino masses are bounded to be 
less than 0.1 eV \cite{Seljak:2006bg}, thus they are very small compared to other 
mass parameters in the theory, and in particular to the electroweak scale 
$v \simeq 246$ GeV. This makes them unobservable in laboratory experiments 
where the relevant energies are much larger, and generically in experiments 
sensitive to electroweak interactions ranging from muon decay to particle 
collisions at LHC. Thus, the question is if light neutrinos have 
further observable interactions beyond their masses. 
This can be answered considering the most general 
effective Lagrangian up to the relevant dimension to be fixed by the available 
experimental accuracy, and fitting it to present data. In general it is enough 
to go to the next order beyond the SM, typically up to dimension six. 
However, the analysis depends, as do the extra operators, on the fields assumed 
to be light and the symmetries preserved. 
Hence, it does depend on the Dirac or Majorana neutrino mass character, 
or more precisely on whether the effective Lagrangian involves or not 
RH neutrinos and on whether lepton number (LN) is conserved or not. As emphasized above, 
neutrino oscillations and then light neutrino masses and mixing are compatible 
with any neutrino mass character, Dirac or Majorana. Hence, we will discuss 
them in turn. The list of dimension six operators preserving the SM gauge 
symmetry and LN can be found in \cite{Buchmuller:1985jz} for operators not involving RH 
neutrino singlets. As a matter of fact, the $SU(3)_C\otimes SU(2)_L\otimes U(1)_Y$ 
gauge symmetry alone implies that all dimension six operators involving only 
SM fields are LNC if they also conserve baryon number \cite{deGouvea:2007xp}. 
This list can be extended to include $\nu_R$ \cite{delAguila:2008ir}. 

Neutrino masses are the only vestige of LNV within the SM if they 
originate from the Weinberg operator in Eq. (\ref{Op5}). 
As the neutrino mass scale is so small, it is appropriate to assume that 
new physics parameterized by dimension six operators 
involving only SM fields or light RH neutrinos is LNC. 
Limits on those operators for LH neutrinos can be found in 
\cite{NSIFits}, being typically at the per cent level in definite models and 
near the expected sensitivity in neutrino oscillation 
experiments. 
Model independent bounds can be one order of magnitude larger. 
They are in general derived assuming only one new operator 
beyond the SM at a time.  
On the other hand, although requiring a precise cancellation (thus at 
least two new dimension six operators, besides the extension of 
the operator set to include light RH neutrinos), there 
is still a (small) window for new interactions with 
observable effects in a near detector at a neutrino factory \cite{delAguila:2009vv}.  


\section{See-saw signatures at LHC}
\label{Majorananeutrinos}

Neutrino oscillations can not decide on the neutrino 
mass character without further interactions. 
However, if the neutrino mass generation mechanism is mediated 
by new particles near the electroweak scale, these and the 
associated mechanism could be established at the LHC. 
This has been often reviewed for see-saw neutrino masses 
\cite{delAguila:2006dx,delAguila:2009bb,delAguila:2008iz,Atre:2009rg}. 
There are three different tree level particle exchanges generating 
the see-saw operator in Eq. (\ref{Op5}). The lepton and 
Higgs doublets can couple to heavy fermions transforming as 
singlets, $N$, or triplets, $\Sigma$, under $SU(2)_L$.  
These are known as see-saw of type I \cite{SeesawI} 
and of type III \cite{SeesawIII}, respectively.    
On the other hand, the two lepton doublets can couple to a scalar 
transforming as their symmetric product. 
This means an $SU(2)_L$ scalar triplet, $\Delta$, 
what we refer to as see-saw of type II \cite{SeesawII}.
If there is no further new physics at the TeV than the 
see-saw mediators, the scalar and fermion triplets will 
be pair produced with electroweak strength at LHC for they 
transform non-trivially under $SU(2)_L\otimes U(1)_Y$. 
As a consequence, their discovery limits for an 
integrated luminosity of 30 fb$^{-1}$ at 14 TeV 
are above half a TeV (see Table~\ref{tab:discoverylimits}). 
\begin{table}[h]
\caption{LHC reach for see-saw mediators with an integrated luminosity 
of 30 fb$^{-1}$ at 14 TeV. For a comparison with other heavy lepton SM 
additions giving multi-lepton signals see 
\cite{AguilarSaavedra:2009ik}.}
\label{tab:discoverylimits}
\begin{tabular}{@{}llllllll@{}}
\hline
See-saw mediator &&& Discovery limit &&& Most significant signals & \\
\hline
\vspace{-0.3cm}
& & & & & & &  \\
Neutrino singlet $N$ (D) &&& Difficult to observe &&& 
$\ell^\pm \ell^\pm \ell^\mp$ &
$\; $\cite{delAguila:2008hw} \\
Neutrino singlet $N$ (M) &&& Difficult to observe &&& 
$\ell^\pm \ell^\pm , \ell^\pm \ell^\pm \ell^\mp$ &
$\; $\cite{delAguila:2007em,delAguila:2008cj,delAguila:2008hw,partonI} \\
Scalar triplet $\Delta$ (NH) &&& 600 GeV &&& 
$\ell^\pm \ell^\pm \ell^\mp , 
\ell^+ \ell^+ \ell^- \ell^-$, fewer $\ell$ &
$\; $\cite{delAguila:2008cj,partonII} \\
Scalar triplet $\Delta$ (IH) &&& 800 GeV &&& 
$\ell^\pm \ell^\pm \ell^\mp, 
\ell^+ \ell^+ \ell^- \ell^-$, fewer $\ell$ &
$\; $\cite{delAguila:2008cj,partonII} \\
Fermion triplet $\Sigma$ (D) &&& 700 GeV &&& 
$\ell^\pm \ell^\pm \ell^\mp, 
\ell^+ \ell^+ \ell^- \ell^-$, up to 6$\ell$ &
$\; $\cite{delAguila:2008hw} \\
Fermion triplet $\Sigma$ (M) &&& 750 GeV &&& 
$\ell^\pm \ell^\pm , \ell^\pm \ell^\pm \ell^\mp$, up to 6$\ell$ &
$\; $\cite{delAguila:2008cj,delAguila:2008hw,partonIII} \\
\hline
\end{tabular}
\end{table}
This is quite different from the heavy neutrino singlet 
case because $N$ can only decay through its mixing with 
the SM leptons $V_{\ell N}$, suppressing the electroweak 
cross-section for single production by the corresponding 
quadratic factor $|V_{\ell N}|^2$. 
Current limits on this mixing, 
$|V_{eN \ (\mu N)}| < 0.05\ (0.03)$ 
\cite{delAguila:2008pw,delAguila:2009bb,LFVlimits}, 
make them difficult to observe. 
The LHC reach and the main signals for the three types of 
see-saw mechanisms are gathered in Table \ref{tab:discoverylimits}. 
Fermion singlets and triplets can be Dirac (D) or Majorana (M), 
in which case events can be LNV as the samples to look at. 
On the other hand, the LHC potential for scalar 
triplets depends on the neutrino mass hierarchy, 
normal (NH) or inverted (IH), because 
this determines their coupling to $\tau$ leptons, which 
do not allow for an efficient scalar mass reconstruction 
\cite{delAguila:2008cj}. 
 
Several comments are in order. The possibility of observing 
LNV events at large hadron colliders due to the production of 
heavy Majorana neutrinos was emphasized long ago \cite{Keung:1983uu}, 
but in the decay of extra charged gauge bosons in left-right models 
\cite{LR}. 
This is still a viable possibility and heavy neutrinos are 
observable at LHC, for the reference luminosity and energy above, 
up to $N$ masses $\sim 2$ TeV for $W_R$ masses 
up to $\sim 4$ TeV \cite{delAguila:2009bb,Ferrari:2000sp}. 
Analogously, 
LHC can probe new neutral gauge boson masses up to 2.5 TeV and $N$ masses 
up to 800 GeV for a leptophobic $Z'\rightarrow NN$ 
\cite{delAguila:2007ua,partonIZP}. 
Hence, heavy neutrinos transforming trivially under the SM 
can be observed at LHC but as products of new interactions. 
Otherwise, the large backgrounds \cite{Sullivan:2008ki,delAguila:2007em} 
for LNV and LNC signals and the small 
mixings with SM leptons make their significance too low for discovery. 
Present limits on these mixings 
follow by comparison with electroweak precision data 
and rare flavor changing processes 
\cite{delAguila:2008pw,delAguila:2009bb,LFVlimits},   
being more stringent in particular for muons due to 
the better determination of the CKM mixing matrix, 
in very good agreement with the SM prediction \cite{Amsler:2008zzb}.  
At any rate, there is some debate about the 
naturalness of so large mixings for a heavy Majorana 
neutrino because 
their size should be given by the see-saw value 
$\sqrt{m_\nu / {\rm TeV}} < 10^{-6}$. 
Although large mixings are parametrically possible \cite{delAguila:2007ap}, 
they still require some (accidental) symmetry \cite{(accidental)symmetry}. 
Finally, although LNV signals, as same sign di-lepton events, have 
much smaller backgrounds than the corresponding LNC ones, 
in this case opposite sign di-lepton events, this 
does not mean that LNC signals are in general less significant 
when samples with larger lepton multiplicities are taken into account 
\cite{delAguila:2009bb}.

Alternatively, even if the interactions directly involved in 
the neutrino mass generation are too suppressed to manifest 
at large colliders, observables with a different physical origin can 
be related to neutrino masses in specific models. 
In such a case LHC could also give new insights on neutrino masses. 
For example, the flavor dependence of the neutralino decay rates 
to charged leptons can provide a determination of the 
corresponding neutrino mixing in specific supersymmetric models 
\cite{Porod:2000hv}, or the observation of new vector-like 
lepton doublets decaying only to $\tau$ leptons can signal  
to a strongly coupled electroweak symmetry breaking sector, 
as recently shown in the context of a holographic composite 
Higgs model
\cite{delAguila:2010vg}. 

\subsection{Lepton flavour violation}

We can learn on neutrino physics from lepton flavor 
changing processes 
\cite{Raidal:2008jk}, 
even if we do not observe new resonances at LHC. 
These transitions require new lepton interactions not banished to very 
high energies to be sizeable. 
This is the case, for instance, of SM extensions 
tackling the hierarchy problem, like supersymmetric, 
Little Higgs, or extra dimensional models \cite{Buras:2009if}. 
In general they predict lepton flavor violating transitions 
at an observable level in forthcoming experiments \cite{Maki:2008zz}. 


\section{Models of neutrino masses}
\label{Models}

Let us comment on models of neutrino masses to conclude. 
In general they do accommodate their tiny scale compared 
with all other masses in the theory, their hierarchical 
splitting, and their mixing matrix quite close to 
tri-bi-maximal mixing \cite{Harrison:2002er}, 
\begin{equation}
  \label{eq:tribimaximalmixing}
U^{\rm TBM}_L = \frac{1}{\sqrt 6}\left(\begin{array}{ccc} 
2 & \sqrt 2 & 0 \\ 
- 1 & \sqrt 2 & - \sqrt 3 \\ 
- 1 & \sqrt 2 & \sqrt 3 \end{array}\right) \ , 
\end{equation}
but there is no simple predictive model of such a pattern, 
on the other hand rather different from that for quarks, 
considered in general compelling. 
Lacking a theory of flavor, 
the challenge is often to make 
this pattern of neutrino masses compatible with 
a given class of models, for instance, constructed to solve 
the hierarchy problem or to unify the gauge interactions
\cite{deMedeirosVarzielas:2006fc}. 
A popular solution is to realize the discrete symmetry 
$A_4$ on the leptonic sector \cite{Ma:2001dn}, for it allows to 
enforce tri-bi-maximal mixing automatically, 
reducing the challenge to prove that the extended 
scalar sector breaks $A_4$ along the correct direction 
and that higher order corrections give small deviations 
from $U^{\rm TBM}_L$ \cite{Altarelli:2010gt}. 

In summary, there is still 
a lot of experimental and theoretical work ahead 
to have a satisfactory understanding of lepton masses and interactions.
In particular, we expect to know if $U_{L 13}$ is different from 0 
and CP violation in the leptonic sector is sizeable 
\cite{Amsler:2008zzb,GonzalezGarcia:2007ib,Oscilaciones,Fits}. 

\begin{acknowledgement}
We are grateful to the Corfu Institute 2009 organizers for their habitual 
kind hospitality, 
and to J.A. Aguilar-Saavedra, M. P\'erez-Victoria, R. Pittau, J. Wudka, 
and M. Zra{\l}ek for earlier collaboration. 
This work has been partially supported by MICINN project 
FPA2006-05294 and Junta de Andaluc\'{\i}a projects FQM 101,
FQM 03048. The work of J.S. (A.C.) has been partially supported 
by a MICINN (ME) Ram\'on y Cajal contract (FPU fellowship). 
\end{acknowledgement}


%
%

\end{document}